\documentclass[conference, a4paper]{IEEEtran}
\IEEEoverridecommandlockouts

\usepackage[font=footnotesize,caption=false]{subfig} 

\usepackage{cite}
\usepackage{amsmath,amssymb,amsfonts}
\usepackage{algorithmic}
\usepackage{graphicx}
\usepackage{textcomp}
\usepackage{xcolor}
\usepackage{booktabs}
\usepackage{glossaries}
\usepackage{subfig}
\usepackage{algorithm}
\usepackage{bbm}
\usepackage{bm}
\usepackage{multirow}
\usepackage{tabularx}
\usepackage{svg}

\usepackage{tikz}
\usetikzlibrary{shapes.arrows}
\usetikzlibrary{plotmarks}
\usetikzlibrary{patterns}
\usetikzlibrary{arrows.meta}
\usetikzlibrary{positioning}
\usetikzlibrary{spy}
\usetikzlibrary{automata}

\usepackage{pgfplots}
\usepgfplotslibrary{colorbrewer}
\usepgfplotslibrary{fillbetween}

\newacronym{fm}{FM}{Foundation Model}
\newacronym{ml}{ML}{Machine Learning}
\newacronym{genai}{GenAI}{Generative AI}
\newacronym{ai}{AI}{Artificial Intelligence}
\newacronym{llm}{LLM}{Large Language Model}
\newacronym{dnn}{DNN}{Deep Neural Network}
\newacronym{aigc}{AIGC}{\gls{ai} Generated Content}
\newacronym{bpp}{BPP}{Bits Per Pixel}
\newacronym{mse}{MSE}{Mean Square Error}
\newacronym{fid}{FID}{Fréchet Inception Distance}
\newacronym{ps}{PS}{Pixel Swapping}
\newacronym{pe}{PE}{Prompt Extension}

\DeclareMathOperator*{\maximize}{maximize}

\def\BibTeX{{\rm B\kern-.05em{\sc i\kern-.025em b}\kern-.08em
    T\kern-.1667em\lower.7ex\hbox{E}\kern-.125emX}}

\title{Generative Network Layer for Communication Systems with Artificial Intelligence}
\author{Mathias Thorsager, Israel Leyva-Mayorga, \IEEEmembership{Member,~IEEE}, Beatriz Soret, \IEEEmembership{Senior Member,~IEEE}, \\and Petar Popovski, \IEEEmembership{Fellow,~IEEE}
\thanks{Mathias Thorsager, Israel Leyva-Mayorga, and Petar Popovski are with the Department of Electronic Systems, Aalborg University, 9220, Aalborg, Denmark (email: mdth@es.aau.dk; ilm@es.aau.dk; petarp@es.aau.dk)} \thanks{Beatriz Soret is with the Telecommunications Research Institute, University of Malaga, 29071, Malaga, Spain and with the Department of Electronic Systems, Aalborg University, 9220, Aalborg, Denmark (email: bsoret@ic.uma.es).}\thanks{This work was partially supported by the Villum Investigator Grant \mbox{``WATER''} from the Velux Foundation, Denmark. The work of B. Soret is partially supported by the Spanish Ministerio de Ciencia, Innovación y Universidades (PID2022-136269OB-I00).}}
\begin{document}
\maketitle

\begin{abstract}
The traditional role of the network layer is the transfer of packet replicas from source to destination through intermediate network nodes. We present a generative network layer that uses \gls{genai} at intermediate or edge network nodes and analyze its impact on the required data rates in the network. We conduct a case study where the \gls{genai}-aided nodes generate images from prompts that consist of substantially compressed latent representations. The results from network flow analyses under image quality constraints show that the generative network layer can achieve an improvement of more than $100$\% in terms of the required data rate.  
\end{abstract}
\glsresetall
\glsresetall
\section{Introduction}

\emph{If it can be predicted, it does not need to be communicated.} This is a straightforward consequence of Shannon's definition of information as a measure of uncertainty. 
However, digital communication systems dominantly operate under the premise that the data created at the source is unpredictable at the destination. Thus, the standard objective of a network is to act as a \emph{dumb pipe} for bits and ensure that a replica of the source packet arrives through the destination, potentially traversing multiple hops. 
Henceforth, the classical role of intermediate 
and edge nodes is to replicate the packet from an input link to one or more output links. This role was generalized by network coding~\cite{Ahlswede2000}, where network nodes can go beyond replication and combine multiple data flows in more general ways.

Prediction can notably improve the performance and resource utilization at the network layer. Caching~\cite{maddahali2014caching}, for instance, relies on prediction of the data that is likely to be relevant for the destination: the source transmits data to an edge node preemptively, which stores this data until requested by the destination. Here the source data does not need to be delivered in real-time; still, both the edge node and the destination receive replicas of the packets generated by the source. 
The next leap in network prediction is exploiting \gls{genai} and its capacity to create synthetic data. Specifically, we propose the use of intermediate and edge nodes equipped with \gls{genai} to create \emph{approximate replicas} of the source packet and reduce the network load.

\begin{figure}[t]
\centering
\includegraphics{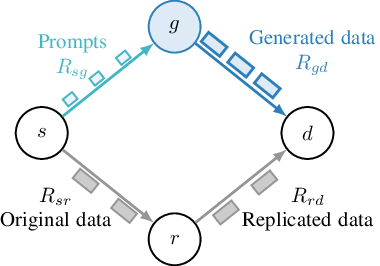}
\vspace{-0.8em}
\caption{Exemplary multi-path network topology with a relay node $r$ that replicates any incoming data and a \gls{genai} node $g$ that generates approximations of the data. $R_{ij}$ is the maximum achievable data rate from node $i$ to $j$.}
\label{fig:examples} \vspace{-0.6cm}
\end{figure}

Fig.~\ref{fig:examples}  shows a simple topology consisting of a source $s$, routers $r$ and $g$, and destination $d$. Node $r$ represents a classical router or relay, while $g$ is a \gls{genai} router that can generate data. Let $R_{ij}$ be the maximum data rate for reliable communication between node $i$ and $j$. Assume a continuous data flow between $s$ and $d$ and a selected route that goes through $r$. Using the max-flow min-cut theorem, the maximal end-to-end data rate is $R_{sd}=\min \{R_{sr}, R_{rd}\}$. This is achieved by having $r$ replicate the received packets to its output link $r-d$ in real time. 
With \gls{genai}, we can reinterpret Fig.~\ref{fig:examples} as follows. If the route traverses through $r$, then $r$ needs to receive the packet from $s$ before creating replica towards $d$. Differently from this, node $g$ can generate an {approximate replica} of the packet that $s$ intends to transmit even without receiving the packet from $s$, based on the prompting from $s$. Let $\mathbf{x}_n$ be the data sent from $s$ at time $n$ and consider the route $s-g-d$, assuming that there is a deadline by which the data should arrive at $g$. The generative network node $g$ can use its GenAI module to create a  sufficient approximation of the data $\hat{\mathbf{x}}_n$ and transmit it to $d$ at a rate $R=R_{gd}$. Here \emph{sufficient} means that $\hat{\mathbf{x}}_n$ is acceptable to $d$ according to some distortion or perception criterion. This approximation is based on a \emph{prompt} sent from $s$ to $g$, which is received by the generative network layer of $g$ to initiate a process of approximate replica generation.

The \gls{genai} model at $g$ can be based on \glspl{fm}, which are \gls{ml} models based on \glspl{dnn}, trained on a broad data set and fine-tuned to specific tasks. Training is chiefly based on self-supervised learning, thus not requiring labelled data. This allows \glspl{fm} to be trained on significantly larger data sets compared to supervised learning. Prompts~\cite{FoundationModels} control the generation of the content by either providing an abstract textual description of the desired content or by providing content of the same data modality for the \gls{genai} to extend. The first case refers to text-to-content models which predominantly consist of \glspl{llm}~\cite{gpt4} and text-to-image or -video models~\cite{Imagen, ImagenVideo}. While the second case also includes \glspl{llm}, many of these models consists of video frame prediction models~\cite{transframer}. To the best of our knowledge, \gls{genai} and \glspl{fm} have not been considered for generative functionality at the network layer. We consider two prompt types for generative network layer: (1) \emph{Implicit} prompting generates new data based on previously received data; when $g$ fails to receive a packet by a certain deadline, it generates approximate replica locally; (2) \emph{Explicit} prompting relies on prompts sent from $s$ to $g$, with a size lower than the actual data size.   

The generation of approximate replicas brings forward the question of quality of reception. In traditional communication systems, reception quality is measured as the distortion between the original and received data due to the use of lossy source coding. The exact distortion metric depends on the particular data type but commonly consist of squared error and Hamming distance~\cite{Rate-Distortion-Perception} for media and sensor data. However, these metrics fail to capture the semantic content, which is the focus of \gls{aigc}. Instead, metrics of perceptual quality of the data must be used depending on the specific data type~\cite{Rate-Distortion-Perception, SemanticMetrics}. In practice, the quality of generated content is evaluated by both distortion and perception metrics.

We present a general model for a \gls{genai}-empowered network layer, outline mechanisms to harness the generative capability of a network node, and analyze the resulting increase of supported data traffic. 
For content based on a set of latent image representations, we discuss the conditions that allow for data approximation and present a mechanism that can be used to gracefully increase or decrease the size of the prompts for image generation. The proposed mechanism compresses an image into a latent representation using a standard library, leading to a notable decrease in data size. By choosing the exact compression method or replacing a specific number of pixels in the generated image with original pixels, we achieve the required prompt size to maximize the flow under specific distortion and \gls{fid} constraints.

\section{System Model}
Fig.~\ref{fig:examples} can be used for a generalized formulation, where we can think of $r$ as a set of nodes, rather than a single one, for multi-hop data transmission for replication. The route via $g$ can provide approximate replicates of the data by generation, and it is often an edge node close to the destination. Let $\mathcal{V}$ be the set of all nodes. Each of the edges $(i,j)$ in the network topology represents an $i-j$ path containing one or more hops between these nodes, which operates under a traditional packet routing model. Therefore, the capacity of each edge $c_{ij}$ represents the capacity of the single- or multi-hop path between a pair of nodes and $r$ is a node in the shortest $s-d$ path that is independent to the $s-g$ and $g-d$ paths. $R_{i,j}$ and $f_{ij}$ are the maximal data rates for reliable communication and the traffic flows between nodes $i$ and $j$, respectively.

The source $s$ attempts to transmit data $\mathcal{X}$ consisting of $N$ individual data packets $\mathbf{x}_n$ each at time $n$ and with an average size $L$. Transmitting $\mathcal{X}$ directly is not feasible as the capacity of the network based on the min cut $\sum_{i\in\{r,g\}}R_{s,i}$ is insufficient. Hence, $s$ transmits a smaller amount of data to $g$ in the form of prompts for intermediate data generation. The two prompting strategies are modeled as follows: 
\begin{itemize}
    \item In explicit prompting, as in~\cite{gpt4, Imagen}, $s$ sends prompts to $g$ at a rate $R=R_p<R_{sg}$, while $R_p$ is sufficient for $g$ to be able to generate acceptable data at a \emph{generation} rate $R_g \geq R_{gd}$. Thus the end-to-end data rate is $R_{sd}=\min \{R_g, R_{gd}\}=R_{gd}$. Explicit prompts are generated from a function $f_\theta(\mathcal{X})$ based on the entirety of the original data, where the parameter $\theta$ determines how the function translates the data into a prompt; 
    \item In implicit prompting, as in~\cite{transframer, Motion-BasedFramePrediction}, $g$ generates $\hat{\mathbf{x}}_n$ based on the data received previously $\mathbf{x}_{n-1},\mathbf{x}_{n-2}, \ldots \mathbf{x}_0$. The prompts consist of the data previously transmitted by the source and the rate of the prompts $R_p$ depends on the amount of data that successfully reaches $g$. For instance, an equivalent prompting rate of $R=R_p<R_{sg}$ is attained by periodically omitting data on the link $s-g$.
\end{itemize}
A middle-ground solution is \emph{hybrid} prompting with a combination of a reduced version of the original data plus the output of the function $f_\theta(\mathcal{X})$. Regardless of the prompt type, the \gls{genai} model is inherently stochastic which means that the same prompt might produce different data when used multiple times. While it is possible to use seeding to guarantee the same output, it is impossible to perfectly predict the output of the GenAI. Hence, prompts cannot be designed to always generate perfect replicas of the original data. Instead, the prompts are designed such that the quality of the generated data is correlated with the size of the prompt. From the sequence of prompts $\mathcal{P}_x=[P_x^{(1)}, P_x^{(2)}, \dotsc,P_x^{(N)}]$, the \gls{genai} model in $g$ uses prompt $P_x^{(n)}$ to generate and transmit the approximated packet $\hat{x}_n$. Considering that the parameters of the \gls{genai} model cannot be changed by the source, the prompt is the only way for the source to control the generated data.

Let $\delta_D(x_{n},g(P_x^{(n)}))$ be the distortion of the $n$-th piece of data, which is a function of the original data $\mathcal{X}$ and the output of the \gls{genai} model $g$ which is given the $n$-th prompt from the set $\mathcal{P}_x$. If the prompts in $\mathcal{P}_x$ are designed properly, the average distortion $\hat{\delta}_D(\mathcal{X},g(P_x)) = \frac{1}{N}\sum_n \delta_D(x_n,g(P_x^{(n)}))$ decreases as the average prompt size $|P_x|$ increases and reaches $0$ when $|P_x|=|x|=L$ since, at this point, the optimal prompt is the data itself. We can redefine the average distortion where $\hat{\delta}_D(L_p)$ is a function of the average size of a prompt $L_p$. Here \emph{average} refers to the expected prompt size required to generate a piece of data with a certain quality, as different types of data may require different prompt sizes to achieve the same distortion. Let $\hat{\delta}_P(g(P_x))$ be the average perceptual quality of the data generated by \gls{genai} model $g$, which is a function of the \gls{genai} model and the prompts used to generate the data. Using the same prompt design philosophy as for the distortion, where lower values from $\hat{\delta}_P(g(P_x))$ indicate higher perceptual quality, the perceptual quality is redefined as $\hat{\delta}_P(L_p)$ which is a function of the average prompt size. 

\section{Problem Definition}

As the distortion and perceptual quality depend on the size of the prompt, there is a trade-off between the possible max-flow gain and the quality of the received data. In a network where $s$ transmits data to $d$ via a \gls{genai}-capable node $g$, the in-flow and out-flow through all nodes except $s$ and $d$ must be equal and the \emph{divergence} of each node is 
\begin{equation}
    y_i=\sum_{j\notin\{s,g,d\}}f_{ij}-\sum_{j\notin\{s,g,d\}}f_{ji} =0.
\end{equation}
However, for node $g$ with the \gls{genai} model we have
\begin{equation}
    y_g=\sum_{j\in\mathcal{V}\setminus g}f_{gj}-\sum_{j\in\mathcal{V}\setminus g}f_{jg} \geq 0,
\end{equation}
 as long as $f_{sg}\geq f_\text{min}$, where $f_\text{min}$ is the minimum flow required by the \gls{genai} algorithm to generate the content. The max-flow gain in a network with \gls{genai} is the increase in flow due to the presence of the \gls{genai} model with respect to the one calculated using the max-flow min-cut theorem $f'_{s,d}$ as
\begin{equation}
    G_\text{flow} = 1+\frac{y_g}{f'_{sd}}.
\end{equation}   
Assuming that prompts are generated at a rate $\lambda$, we optimized: 
\begin{IEEEeqnarray*}{Cll}
\maximize_{L_p, \lambda}\quad & y_g - y_g w \hat{\delta}_m(L_p) \qquad& m\in\{D,P\}\\
\text{subject to } 
&f_{si} \leq c_{si} & \forall i \in \mathcal{V}\setminus s\\
&f_{id} \leq c_{id} & \forall i \in \mathcal{V}\setminus d\\
&\lambda L_p = f_{sg} \geq f_{min}\\
&\lambda L = f_{gd}.
\end{IEEEeqnarray*}
where $w$ is a scalar that determines the importance of the quality of the generated content $\hat{\delta}_m(L_p)$ and $L_p\in\left(1,L\right]$ is the average size of the prompt for a given strategy. By taking the flow constraints $f_{sg}\leq c_{sg}$ and $f_{gd}\leq c_{gd}$, it is straightforward to find the optimal value as

\begin{equation} \label{eq:dataGenRate}
    \lambda^*=\min\left(\frac{c_{sg}}{L_p},\frac{c_{gd}}{L} \right).
\end{equation}

\section{Case Study: Image Generation}

To investigate the benefits of network content generation, we conduct a case study on the transmission and generation of images. For the image generation we use an implementation of explicit prompting where the prompts consist of latent representations of images. For this, we relate the rate-distortion and rate-perception functions $\hat{\delta}_D(L_p)$ and $\hat{\delta}_P(L_p)$ to quality metrics related to image generation. In particular, we use \gls{mse} as the distortion metric, and \gls{fid}~\cite{FID} as the perceptual quality metric. \gls{fid} is a perceptual quality measure which calculates the difference between two Gaussian models fitted to two data sets: real-world and generated images. This means that it cannot evaluate the perceptual quality of individual images, but given a data set of generated images it can measure how well these resemble real images. Since the quality measure is dependent on a data set of real images, the resulting \gls{fid} score for a set of generated images cannot be interpreted as the actual \emph{globally true} quality but a relative score for the particular data set of control images. However, by keeping the control data set constant, a comparison of the relative score between different data sets of generated images is still meaningful. The rate is measured as the size of the prompts used to generate the images. To ensure the prompt size is invariant to changes in image dimensions, it is defined as the average \gls{bpp} of the latents calculated as the size in bits of the latents divided by the number of pixels in the original images. Both the conversion of image to latent and the generation on an image based on the latent is performed using the models presented in \cite{HiFiC}. We use three \gls{genai} models and corresponding compression models, i.e., we can compress images into latents of three different \gls{bpp} values and using these generative approximations. Moreover, we present two approaches for extending the size of the latents: (1) \gls{pe} of \gls{genai} models that can take latents of any arbitrary \gls{bpp} value. (2) \gls{ps}, where parts of the original image are transmitted along with the latent and replaced at the receiver. For both methods, the rate-quality functions are found using a process of curve-fitting based on a set of prompt size and distortion or perceptual quality value pairs. Since it is not known which functions best approximate the data beforehand, the first step of the curve-fitting is a visual inspection of the data to estimate the type of the function (e.g., polynomial, exponential). With a function type selected, the curve-fitting itself consists of the maximisation of the $r^2$ function.

\subsubsection{Prompt Extension}
We find the rate-perception function $\hat{\delta}_P(L_p)$ for \gls{pe} using curve-fitting based on a set of prompt size and perceptual quality pairs. The prompt size is calculated as the average \gls{bpp} of latents using the three compression models from \cite{HiFiC} on a dataset of images. The perceptual quality is likewise calculated as an average quality over the dataset for the three \gls{genai} models using the FID metric. Besides these three data points, we use the \gls{bpp} of the true images to define the upper bound of the function where the FID is 0. While the resulting function of the curve-fitting using these four points is defined at a prompt size of 0 \gls{bpp}, it does not make sense to extend the function to this point. Instead the lower bound has to be defined somewhere between 0 \gls{bpp} and the \gls{bpp} of the \gls{genai} model with the smallest latents. For this work, the lower bound is set at the \gls{bpp} of said \gls{genai} model. 

To enable the use of the \gls{fid} values as weights in the optimization problem, we normalize the \gls{fid} values using an assumed maximum and minimum \gls{fid} value. From an initial curve-fitting of the \gls{pe}, we find the output of the rate-perception function when the prompt size is set to 0. This is assumed as the maximum \gls{fid} for the \gls{genai}-based image generation used in this paper. The minimum \gls{fid} is set as 0 indicating the transmission of the true images. With these, the normalized rate-perception function is found using curve-fitting but with the normalized \gls{fid} values.

\subsubsection{Pixel Swapping}
\Gls{ps} refers to replacing parts of the generated image with the true image. It is a practical approach to generate prompts of any size
and can be combined with any of the three \gls{genai} models described above to achieve different rate-distortion and -perceptual quality trade-offs. Specifically, we use curve-fitting based on a set of \gls{bpp} and distortion or perceptual quality pairs to obtain three continuous rate-distortion and -perception functions; these are referred to as the low, medium, and high \gls{ps} functions. The average \gls{bpp} of the combined images $L_c$ is calculated by extending the prompt with a fraction $\gamma$ of the true images as 
\begin{equation}\label{eq:combPrompt}
    L_c = L_p + \gamma L, 
\end{equation}
where each pixel is chosen uniformly randomly. For each of the image combinations using different amounts of replaced pixels, the perceptual quality is calculated using \gls{fid} the same way as for the \gls{pe} including the normalization step. The distortion is calculated as the normalized average \gls{mse} of the combined image and the true image. The average \gls{mse} is normalized based on the maximum and minimum \gls{mse} -- the maximum \gls{mse} is the \gls{mse} between the true image and a color inverted version of the image, and the minimum \gls{mse} is defined as 0 (the \gls{mse} of two copies of the true image).

\section{Results}
The rate-quality functions are fitted to a generated approximation of a dataset consisting of 2048 randomly chosen images from the COCO2017~\cite{CoCo} training dataset. Furthermore, the \gls{genai} based transmission scheme is compared against a baseline image compression scheme using the JPEG compression standard. While JPEG does not compress images into latents which are generated into images at the \gls{genai} equipped node $g$, we obtain "prompts" of different sizes by adjusting the quality setting in the JPEG compression which determines the compression ratio and therefore also the quality of the resulting compressed images. 

\subsection{Curve-Fitting}

\begin{figure}
    \centering
    \includegraphics[width=1\linewidth]{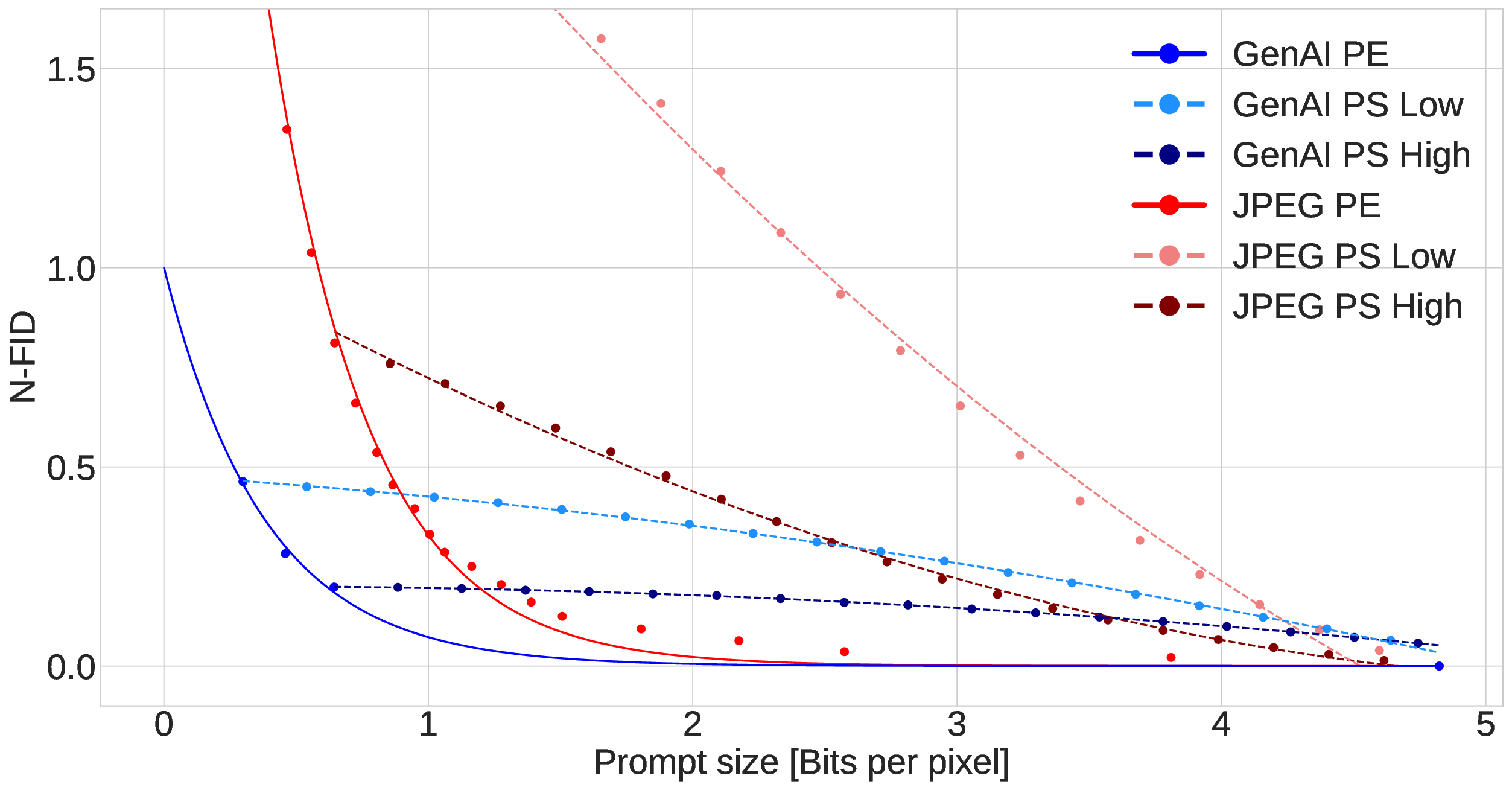}
    \caption{Rate-Perception functions for the Prompt Extension(\gls{pe}) and Pixel Swapping (\gls{ps}) approaches for the \gls{genai} and JPEG compression schemes. The dots indicate the measured data points and the lines are the curves fitted to these data points.}
    \label{fig:FIDvBPP}
\end{figure}

\begin{figure}
    \centering
    \includegraphics[width=1\linewidth]{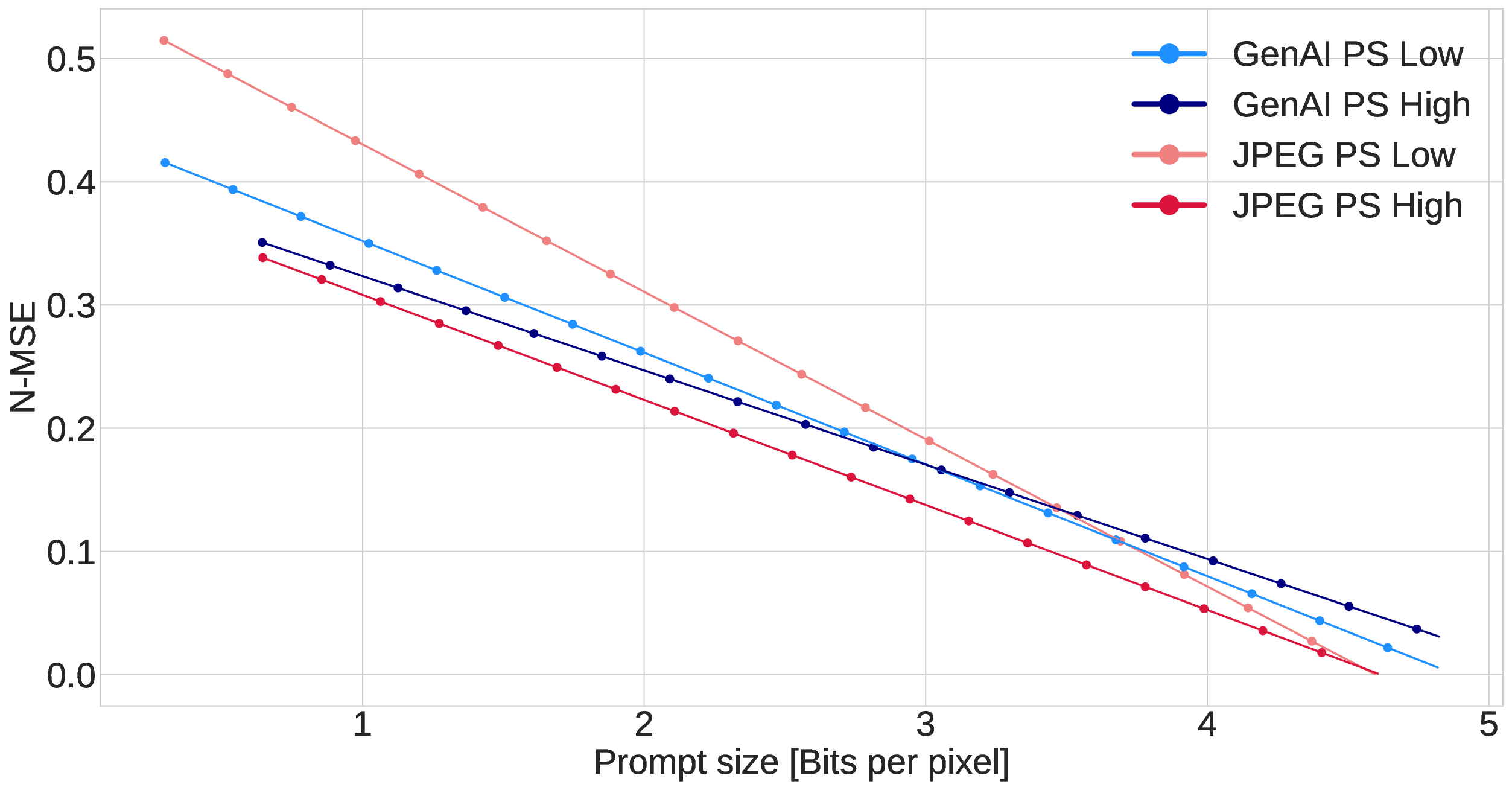}
    \caption{Rate-Distortion functions for the Pixel Swapping (\gls{ps}) approaches for the \gls{genai} and JPEG compression schemes. The dots indicate the measured data points and the lines are the curves fitted to these data points.}
    \label{fig:MSEvBPP}
\end{figure}

The rate-perception functions from the curve fitting are shown in Fig.~\ref{fig:FIDvBPP} for both the \gls{genai} and JPEG compression based image transmission schemes. The figure shows that \gls{pe} is better than \gls{ps} for all prompt sizes for both the \gls{genai} and JPEG compression schemes. While this means that it would always be better to generate a latent of the given prompt size rather than using \gls{ps}, the \gls{pe} approach is, at present, only a theoretical result. The \gls{genai} based scheme presented in \cite{HiFiC} does not allow for the use of arbitrary latent sizes as each model is specifically trained with a particular compression model which cannot dynamically change the size of the resulting latents. However, we expect that this restriction is only temporary and later models will overcome this shortcoming. As such, even though \gls{ps} is worse than \gls{pe}, it is the only practically available approach. Between the two schemes, we see that for \gls{pe}, the \gls{genai} based scheme shows better or equal perceptual quality as the JPEG compression scheme for all prompt sizes. However, for \gls{ps}, the \gls{genai} based scheme only shows a better quality for prompt sizes lower than 3.6~\gls{bpp}. For the larger prompt sizes, the JPEG compression does show slightly better perceptual quality.

Fig.~\ref{fig:MSEvBPP} compares the \gls{ps} rate-distortion functions for the \gls{genai} and JPEG compression based schemes. For the majority prompt sizes, the JPEG compression shows a better distortion than the \gls{genai} based scheme. This does, however, make sense as the JPEG compression is less costly in terms of prompt size when it is not required to transmit the entire latent (the compressed image in the case of JPEG compression). As such, despite the two schemes being close in distortion when no pixels are swapped for the large latent size, the prompt size of the \gls{genai} scheme increases faster than for the JPEG. Only for the smallest prompt sizes does the \gls{genai} based scheme show better distortion than the JPEG compression.

\subsection{Flow Optimization}

\begin{figure}[t]
    \centering
    \subfloat[Optimal prompt size\label{subfig:Optim_QualityWeight}]{\includegraphics[width=0.5\linewidth]{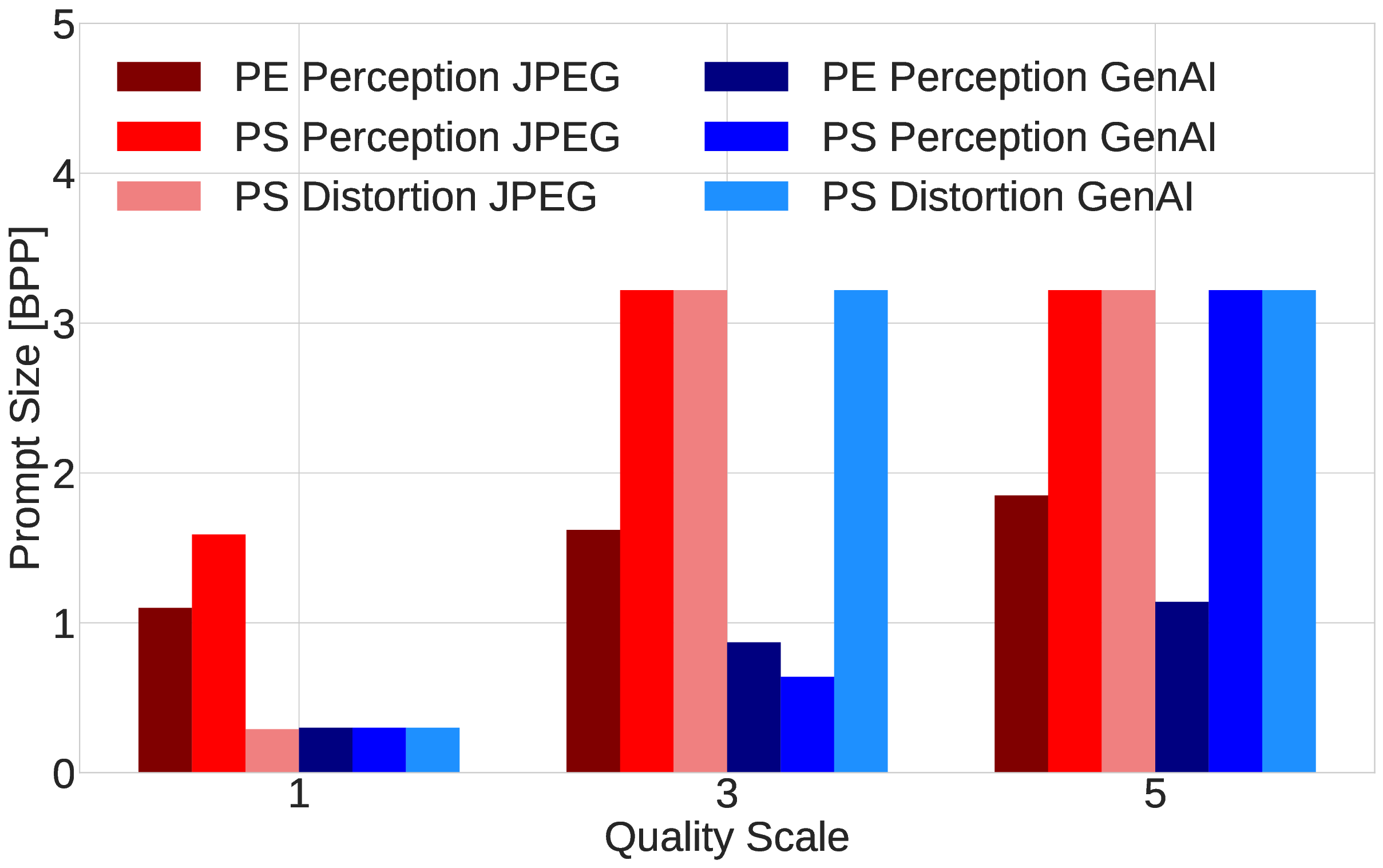}}
    \subfloat[Flow gain $G_\text{flow}$\label{subfig:Optim_G_QualityWeight}]{\includegraphics[width=0.5\linewidth]{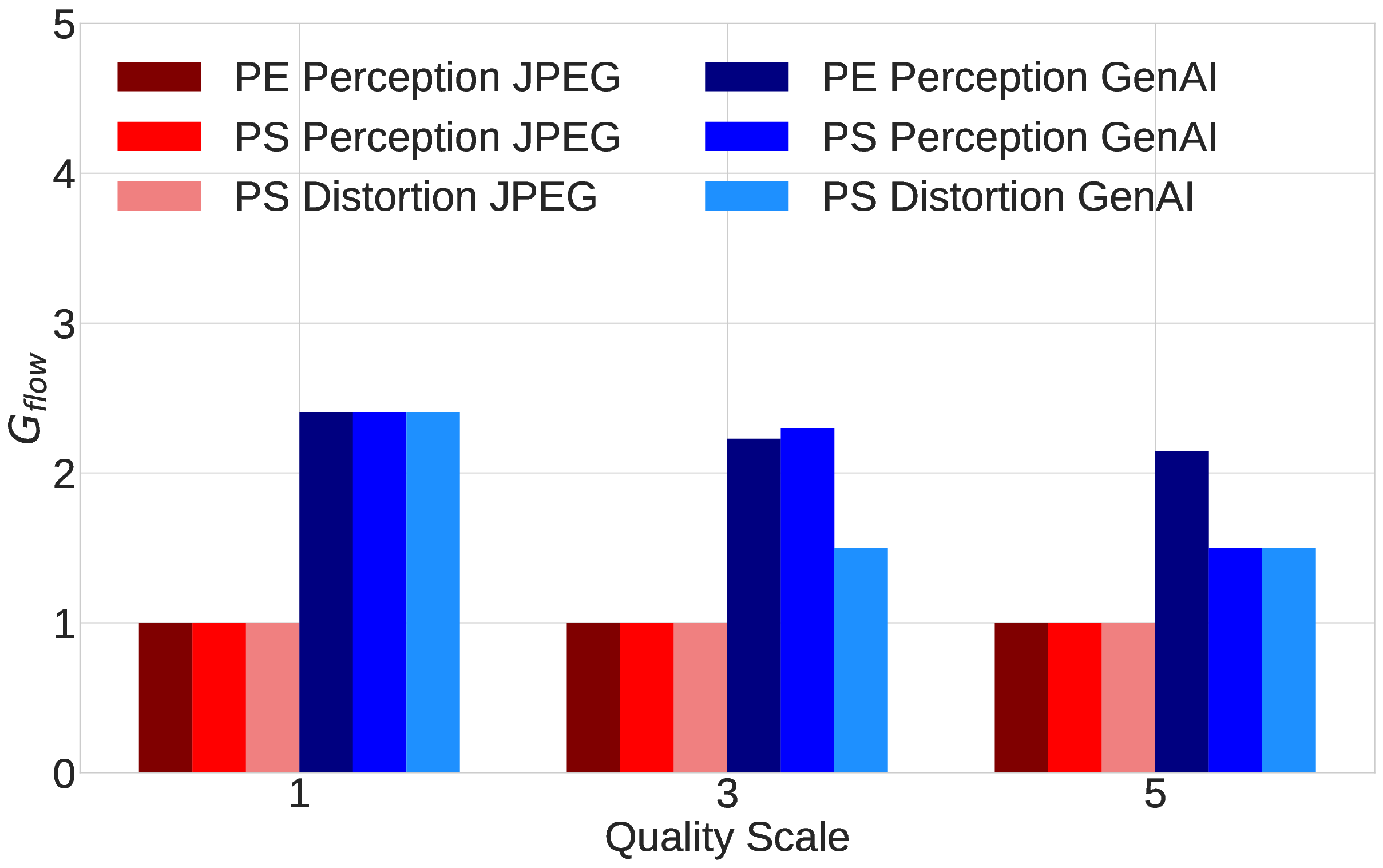}}
    \caption{Results of the optimization problem with $c_{sg}=3.184$~\gls{bpp}: (a) The Optimal prompt size given rate-distortion and -perception functions for the two proposed approaches. (b) The $G_\text{flow}$ calculated from the optimal prompt size and the size of the content after node $g$.}
    \label{fig:optim_Res} \vspace{-0.5cm}
\end{figure}

Fig.~\ref{subfig:Optim_QualityWeight} shows the results of the optimization problem when increasing the quality scale $w$. At the low quality scale, all but the rate-perception functions for the JPEG compression find the optimal solution by transmitting as little information as possible. As seen in Fig.~\ref{fig:FIDvBPP}, the rate-perception functions for the JPEG compression do not scale well at the smaller prompt sizes and are therefore required to transmit more to maintain a better quality despite the low quality scale. As the quality scale increases, the \gls{ps} functions shift from transmitting only the latent, to transmitting as many original pixels as the capacity $c_{sg}$ allows. The optimization functions for the three \gls{genai} compression models are always convex which, means that the optimal prompt size is found at one of the functions' bounds. That is, depending the available link rate, the optimal solution for \gls{ps} is either not sending original pixels or sending as many original pixels as possible.

In spite of the convex optimization functions of the \gls{genai} based \gls{ps} approaches, \gls{ps} still allows for other optimal solutions to the optimization problem than just the original three latent prompt sizes. We see that for higher quality scales, the optimal solution is to transmit at the capacity which is only possible due to \gls{ps}. As the capacity changes, only the upper bound of the optimization problem is affected. As such, allowing the rate-quality function to extend to this bound is shown to be better. The convex optimization function also explains why the prompt size is lower for the \gls{ps} based rate perception function than \gls{pe} for the \gls{genai} scheme with the middle quality weight. Intuitively, a lower prompt size indicates a better result as it implicates a higher achieved flow. However, the prompt size alone does not show the actual value from the optimization problem. Therefore, even though \gls{ps} resulted in a lower prompt size, it did so with a lower optimization problem value than \gls{pe}. However, when we compare the two \gls{pe} approaches in the middle and large quality scales, the fact that the \gls{genai} finds optimal solutions at lower prompt sizes than the JPEG scheme does indicate a higher performance. As shown in Fig.~\ref{fig:FIDvBPP}, the \gls{genai} based \gls{pe} approach manages a higher perceptual quality for all prompt sizes. This translates to a better optimization problem result regardless of the quality scale.

While the optimal prompt size for the \gls{pe} based \gls{genai} scheme does show a significant improvement over the JPEG compression, up to more than a 100\% gain in data flow under the low quality constraint, the main benefit is seen in Fig.~\ref{subfig:Optim_G_QualityWeight}. Due to the fact that the JPEG compression does not alter any data at node $g$ (it simply acts as a replicator), it sees no max flow gain regardless of quality scale or capacity. As such,
we see that the \gls{pe} approach maintains more than a 100\% increase in max flow gain compared to the JPEG compression scheme for all three quality constraint scales. Furthermore, it shows that the for the given capacity of 3.184 \gls{bpp}, the \gls{ps} approach maintains a max flow gain of more than 50\% for the three quality constraint scales.

\section{Conclusion}
We presented a model for a generative network layer, where \gls{genai} models are used in intermediate network nodes to generate data approximations from prompts of varying sizes. Through an example of generative image transmission, we compared the performance of the proposed \gls{genai}-empowered network layer with JPEG image compression. The results show that the \gls{genai}-based scheme offers a higher perceptual quality at lower prompt sizes. This results in a higher effective data flow through the network, under given perceptual quality constraints. Furthermore, the proposed prompt size extension approaches improve the optimized values of the data flows.

\bibliographystyle{IEEEtran}
\bibliography{sources}
\end{document}